\documentclass[12pt,preprint]{aastex} 

\shortauthors{Groh, Hillier \& Damineli}
\shorttitle{Rotation of AG Car}

\begin{document}

\title{AG Carinae: a Luminous Blue Variable with a high rotational velocity\footnote{Based on
observations made at the 1.52\,m ESO telescope at La Silla, Chile and at the 8\,m VLT/ESO
at Paranal, Chile}}

\author{J. H. Groh\altaffilmark{1,2}}
\email{groh@astro.iag.usp.br}
\author{D. J. Hillier\altaffilmark{2}}
\email{hillier@pitt.edu}
\author{A. Damineli\altaffilmark{1}}
\email{damineli@astro.iag.usp.br}
\altaffiltext{1}{Instituto de Astronomia, Geof\'{\i}sica e Ci\^encias 
Atmosf\'ericas, Universidade de S\~ao Paulo, Rua do Mat\~ao 1226, Cidade 
Universit\'aria, 05508-900, S\~ao Paulo, SP, Brasil}
\altaffiltext{2}{Department of Physics and Astronomy, University of Pittsburgh,
3941 O'Hara Street, Pittsburgh, PA, 15260, USA}

\begin{abstract}

We report the detection of broad absorptions due to \ion{Si}{4} $\lambda\lambda$4088--4116 in the
Luminous Blue Variable (LBV) AG Carinae during its last hot phase (2001-2003). Our NLTE spectral
analysis, with the radiative transfer code CMFGEN, revealed the photospheric nature of these lines
predicting, however, much narrower and deeper absorption profiles than observed. Using a
recently-developed code to compute synthetic spectra in 2D geometry allowing for the effects of
rotation, we could match the broad absorptions with a high projected rotational velocity of 190 $\pm$
30\,km\,s$^{-1}$ on 2001 April. Analysis of spectra obtained on 2002 March and 2003 January, when the
star was cooling, yielded to a projected rotational velocity of 110\,$\pm$\,10\,km\,s$^{-1}$ and
85\,$\pm$\,10\,km\,s$^{-1}$, respectively. The derived rotational velocities are proportional to $R_{\star}^{-1}$, as
expected from angular momentum conservation. We discuss the effects of such high rotation on the
spectral analysis of AG Car, and on the wind terminal velocity. Our results show direct
spectroscopic evidence, for the first time, that a LBV may rotate at a significant fraction of its
break-up velocity. Thus, AG Car (and possibly other LBVs) is indeed close to the $\Gamma\Omega$ limit,
as predicted by theoretical studies of LBVs.

\end{abstract}


\section{\label{intro}Introduction}

Massive stars have a tremendous impact on both the chemical and dynamical evolution of a galaxy.
They input energy and momentum, and eject their nucleosynthesis products into the interstellar medium
\citep{fhy03}. A detailed knowledge of massive star evolution is thus crucial to understanding
their host galaxies, especially in the early Universe when massive stars are dominant
\citep{mc05}.

Over the last decade, significant efforts have been made to include the effects of rotation
consistently in the evolutionary models for massive stars \citep{mm_rot_paper1}. It has been
found that the effects of rotation are far from negligible, especially for the most massive
stars with high initial rotational velocity. As a consequence of rotation, the evolutionary tracks on the HR
diagram, and the duration of the evolutionary stages (main sequence, Luminous Blue Variable --
hereafter LBV, and Wolf-Rayet phases), are substantially modified. For a recent
review we refer to \citet{maeder_araa00}.

Rotation is also expected to play an important role in both LBV variability and LBV
eruptions \citep{langer99,mm_omega00,alm04}. Even before reaching the Eddington limit,
these stars may lose large amounts of material if they have a significant rotational
velocity compared to their critical rotational velocity. Stellar rotation, during the
LBV phase, can have an impact on the surrounding circumstellar medium, shaping their
ejected nebulae with the characteristic bipolar geometry found around many of these stars
\citep{hd94,md01,do02}.

In this Letter we present high-resolution, high S/N optical spectra of the LBV
AG Carinae. AG Car has pronounced S Dor type variations, changing its effective temperature from  
9,000 to 25,000 K \citep{leitherer94,stahl01}. AG Car is considered the prototype of the LBV class
and has been extensively
observed, from UV to radio wavelengths, with photometric, spectroscopic and polarimetric
techniques over several decades. It also has a bipolar nebula which is seen at various
wavelengths \citep{nota92,voors00,dw02}  that was ejected approximately 10$^4$ years ago
\citep{lamers01}.

Here we report the detection of broad absorptions associated with the high-excitation lines of
\ion{Si}{4} $\lambda\lambda$4088--4116 in AG Car during its last hot phase (2001-2003). These data are
described in Sec. \ref{obs}, while in Sec. \ref{modres} we present the results of our NLTE spectral
modeling. Our modeling reveals the photospheric nature of the \ion{Si}{4} lines and allow us to derive,
for the first time, the rotational velocity ($v_{rot}$) of AG Car and how it varies during the S
Dor cycle. We show that $v_{rot}$ is high compared to its critical rotational velocity ($v_{crit}$). In
Sec. \ref{disc} we discuss some of the effects expected to be present in fast-rotating stars, such as
the reduction in the wind terminal velocity and the presence of asymmetric winds. The high
$v_{rot}$ derived has important implications not only for AG Car and its wind, but for the general
understanding of the LBV phase and the subsequent evolutionary stages of a massive star.

\section{\label{obs}Observations}

The spectroscopic observations of AG Car analyzed in this work were obtained as part of
an observational program to follow the star over a complete S Doradus cycle with a dense
time sampling and at high resolution. The results provided by this campaign will be
published in a forthcoming paper (Groh et al. in prep.).

During 2001 and 2002, we collected AG Car spectra at the 1.52\,m telescope of the European
Southern Observatory (ESO) in La Silla, Chile, with the fiber-fed, bench-mounted echelle
spectrograph FEROS \citep{kaufer99}. To reduce the data we utilized  an automatic pipeline
reduction system; further details are given in \citet{stahl99}. The configuration used on the FEROS spectrograph
provided a resolution of 48,000 over the region 3600--9200 {\AA}, with increasing
signal-to-noise ratio (S/N) from blue to red. We observed AG Car with FEROS on 2001 April
12 and 2002 March 17.

In addition to the FEROS spectra, we retrieved spectra of AG Car from the ESO public-data
archive\footnote{\url{http://archive.eso.org}}. They were obtained on 2003 January 11 with the UVES
spectrograph attached to the 8m VLT-Kueyen at ESO, under ESO Director Discretionary Time Program
266.D-5655 \citep{bagnulo03}. The resolution provided by the UVES spectrograph was 80,000 in the
wavelength range 3000--10400 {\AA}, with an exceptional S/N of about 400. As with the FEROS spectra,
we utilized the pipeline-processed data provided online. Table \ref{obstable} summarizes the
high-resolution spectroscopic observations utilized in this Letter. 

\begin{figure}
\plotone{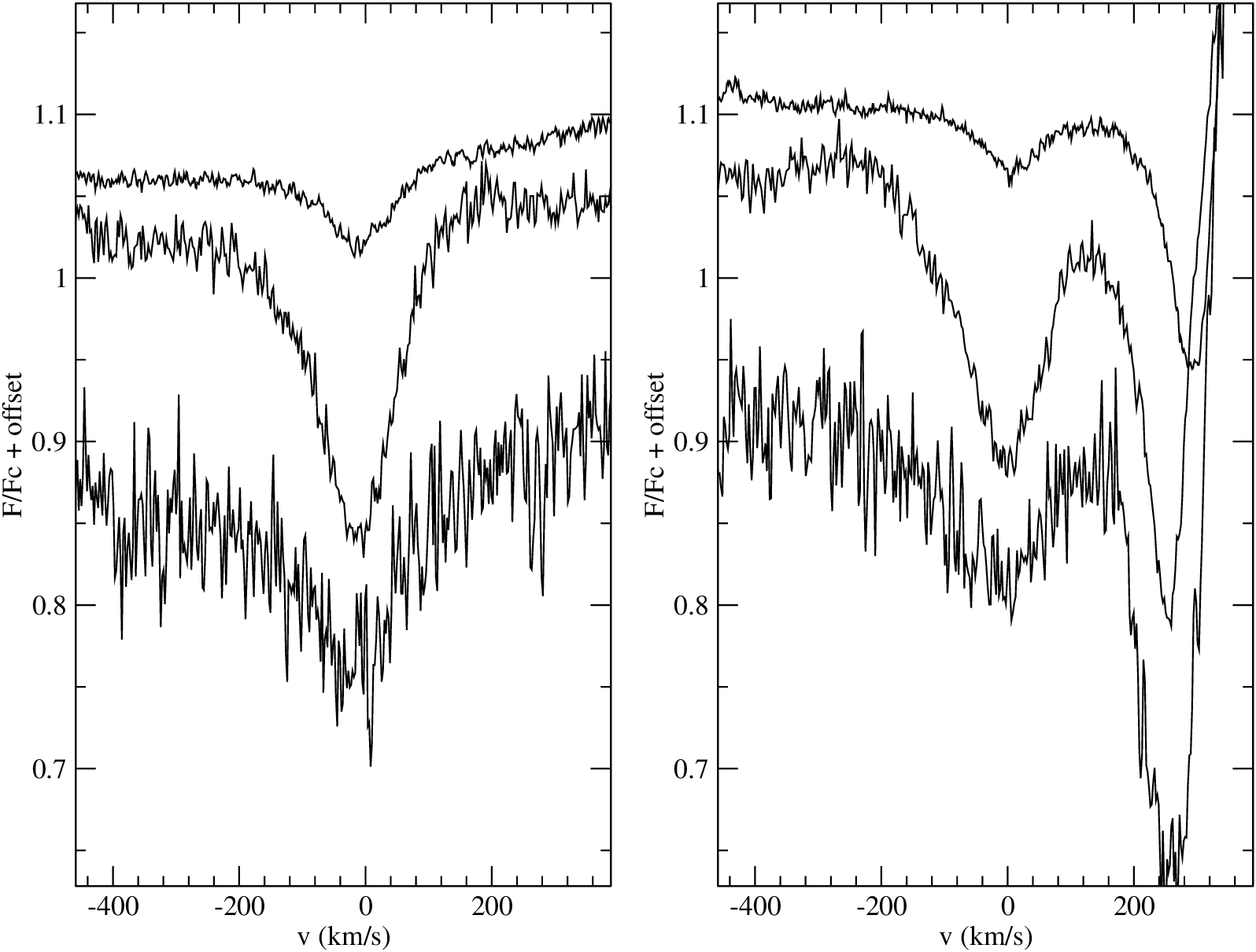}
\caption{High resolution, continuum-normalized spectra of AG Car, centered on Si
IV 4088.86 {\AA} (left panel) and on \ion{Si}{4} 4116.10 {\AA} (right panel). The
spectra were offset for the sake of clarity.  On both panels, we show from
top to bottom data from 2003 January (+0.05 offset), 2002 March (no offset) and
2001 April (-0.15 offset). Significant changes in the strength of the \ion{Si}{4} lines, related
to changes in the effective temperature of the star, can be seen. \label{agcobs}}
\end{figure}

In Figure \ref{agcobs} we show (in velocity space) the absorption profiles of \ion{Si}{4}
$\lambda\lambda$ 4088--4116 detected in  AG Car during its last hot phase. From bottom to top,
we show the 2001 April, 2002 March and 2003 January spectra. The spectra were corrected to the 
heliocentric frame and the systemic velocity of 10\,km\,s$^{-1}$ \citep{stahl01} was subtracted. The same broad \ion{Si}{4}
absorption profiles can also be seen in high-resolution FEROS spectra that we have taken in 2001
January, 2001 June and 2002 July, that are omitted here due to their low S/N, and in
lower resolution spectra. Hence, we conclude that the \ion{Si}{4} broad absorptions are present
during the whole hot phase (i.\,e., they are not a transient feature).

\section{\label{modres}Modeling and results}

To analyze the observational spectra of AG Carinae we used CMFGEN \citep{hm98}. CMFGEN assumes an
outflow in steady-state,  spherical symmetry and treats line and continuum formation in NLTE. Each
model is specified by the hydrostatic radius $R_\star$, the luminosity $L$, the mass-loss rate
$\dot{M}$, the volume filling-factor $f$ and the chemical abundances $Z_i$ of the included species. The
code does not solve the momentum equation of the wind, so a velocity law must be specified. The
velocity structure is described by a beta-law modified at depth to match an hydrostatic structure at
$R_\star$ (typically at a Rosseland depth of 100 in our models), so we also have to input the wind
terminal velocity $v_{\infty}$ and the steepness of the velocity law $\beta$. We assumed a constant
turbulent velocity of 20\,km\,s$^{-1}$. A higher turbulent velocity would give a detectable redshift,
which is not observed, of the emission lines with high opacity, such the stronger lines of \ion{H}{1}
and \ion{He}{1}.

While CMFGEN does not handle directly the effects of rotation in the radiative transfer, we accounted
for its influence on the observed spectrum  following the procedures described by \citet{bh05}. For
each CMFGEN model we computed the emerging spectrum in 2D geometry viewed from pole (0 degrees, which
was equivalent to a non-rotating model), 30 degrees and equator (90 degrees), for a given value of
$v_{rot}$. As our goal is only to show that the broad absorptions are photospheric and broadened by
rotation, we did not explore the effects of an equatorial or polar density enhancement on the line
profiles. As shown by \citet{bh05}, the first order impact will be on the relative line strengths,
which would firstly affect the derivation of the temperature and/or mass-loss rate. These effects will
be investigated in a forthcoming paper.

\begin{figure}
\plotone{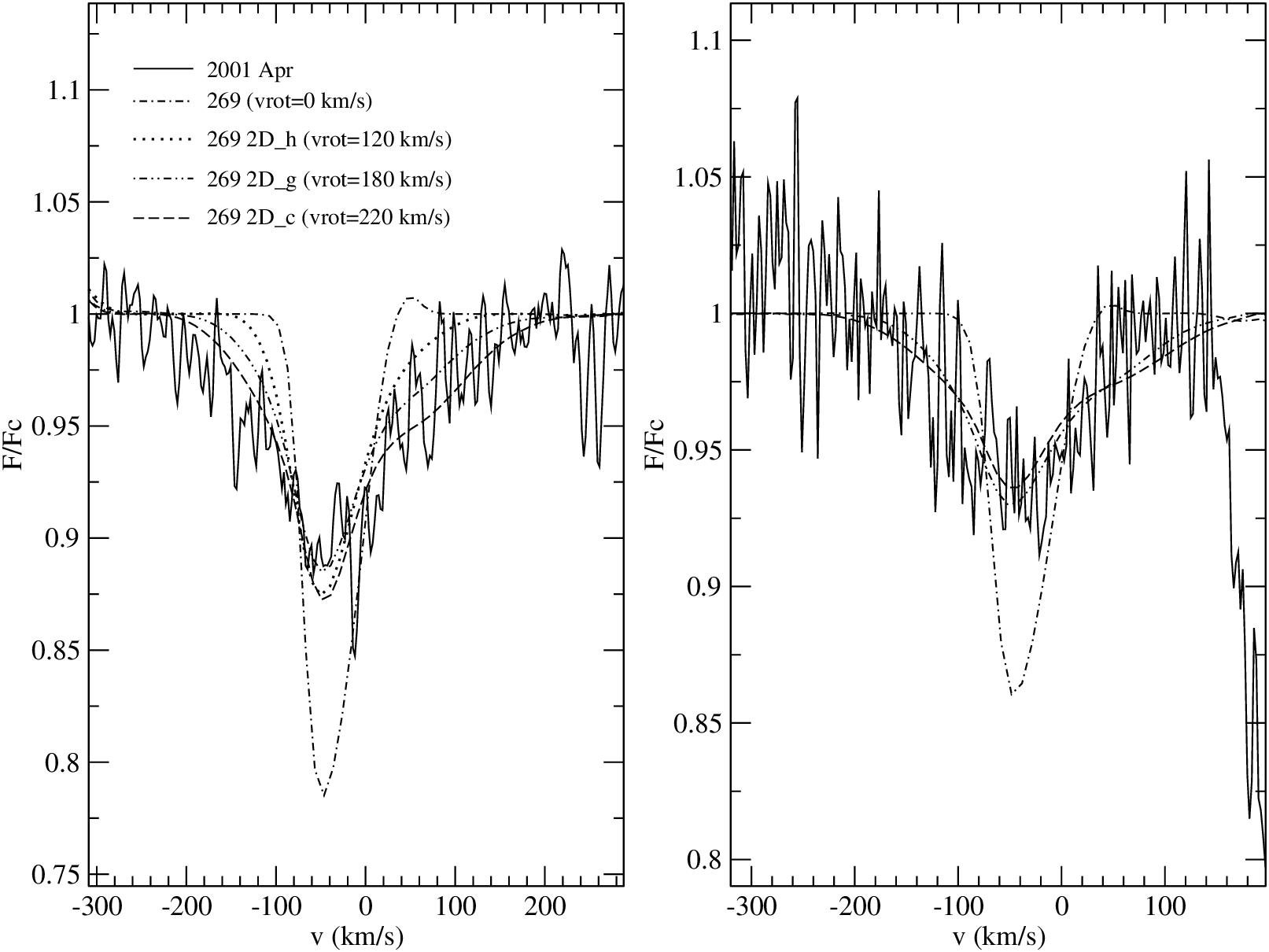}
\caption{High-resolution observed absorption profiles (full line) at
\ion{Si}{4} 4088
(left panel) and \ion{Si}{4} 4116 (right panel). A set of CMFGEN models, computed without
rotation (dot-dashed),  and with a projected rotational velocity  of 120\,km\,s$^{-1}$ (dotted 
line), 180\,km\,s$^{-1}$ (double dot-dashed) and 220\,km\,s$^{-1}$ (dashed line). For the sake of
clarity, at the right panel we only show the non-rotating, 180\,km\,s$^{-1}$ and 220\,km\,s$^{-1}$
models. In spite of the errors due to the low S/N, we can clearly rule out a rotational velocity
lower than $\sim$150\,km\,s$^{-1}$ for AG Car on this epoch.\label{agc01}}
\end{figure}

In Figure \ref{agc01} we show the absorption profiles of \ion{Si}{4} $\lambda\lambda$ 4088--4116
observed on  2001 April compared with our best model. Based on the spectroscopic analysis, we obtained
for this epoch L=1.0$\times10^{6}$\,L$_\odot$,  R$_\star$=56\,R$_{\odot}$, T${_\star}$=24,340\,K,
v$_{\infty}$=105\,km\,s$^{-1}$, $\dot{M}$=2.3$\times10^{-5}$\,M$_\odot$\,yr$^{-1}$, and f=0.15. The
non-rotating model predicts significantly narrower and deeper absorption profiles for
both \ion{Si}{4} lines than the observations. As these \ion{Si}{4} lines are formed deep in
the wind, the broadening effects due to the velocity field are negligible. We needed
$v_{rot}$\,sin\,i=190$\pm$30\,km\,s$^{-1}$ to reproduce the observed lines' shape and width, as can be
seen analyzing the fits of model 269\_2D\_g and 269\_2D\_c. In
spite of the noise of the data, we can clearly rule out a lower value of $v_{rot}$ (e.\,g. a
non-rotating model or model 269\_2D\_h with $v_{rot}$\,sin\,i=120\,km\,s$^{-1}$).

To derive $v_{crit}$, and the surface escape velocity ($v_{esc}$), we adopted the mass of AG Car as
$M=35\,M_{\odot}$ \citep{sc94,vink02}, which should be regarded as a lower limit (see discussion in Sec.
4 of Vink \& de Koter 2002). Changes in $M$ will affect the analysis, since $v_{crit}$ and $v_{esc}$ are
both proportional to $M^{0.5}$, and the Eddington parameter $\Gamma$ depends on $M^{-1}$. During 2001
April we calculated $v_{crit}$=220\,km\,s$^{-1}$, following \citet{mm_omega00}, for a derived value of
$\Gamma$=0.7. Assuming that we are seeing AG Car edge-on, the derived $v_{rot}$=190\,km\,s$^{-1}$
amounts to 0.86\,$v_{crit}$. This value is exceptionally high and it is a lower limit, since higher
values will be inferred for lower inclination angles $i$. For example, if  $i=60^o$, we would have
$v_{rot}$=220\,km\,s$^{-1}$ and $v_{rot}/v_{crit}$=1, i.\,e., the star would rotate nearly at
$v_{crit}$. 

The precise determination of $v_{rot}/v_{crit}$ is beyond the scope of this Letter, since it relies on
the accuracy of $M$, $L$, distance, reddening law, and $i$. However, for any reasonable change in
one or more of these parameters, a significant value of $v_{rot}/v_{crit}$ would still be reached.
Hence, the rotationally-broadened photospheric absorptions reported in this Letter make it implausible
that AG Car is seen pole-on or from lower inclination angles, as $v_{rot}$ would be arbitrarily high.
Moreover, the kinematics and dynamics of its bipolar nebula show extremely low projected velocities, in
the range $\pm$50\,km\,s$^{-1}$ \citep{nota92}, which is consistent with a two-lobe polar ejection
viewed nearly from the equator. 

It is important to point out that our models with rotation predict undetectable changes 
on the computed profiles of \ion{H}{1}, \ion{He}{1} lines, and other strong lines on AG Car's
spectrum. As they are formed far from the photosphere, the influence of rotation is very low,
due to the $r^{-1}$ effect (conservation of angular momentum).

\begin{figure}
\plotone{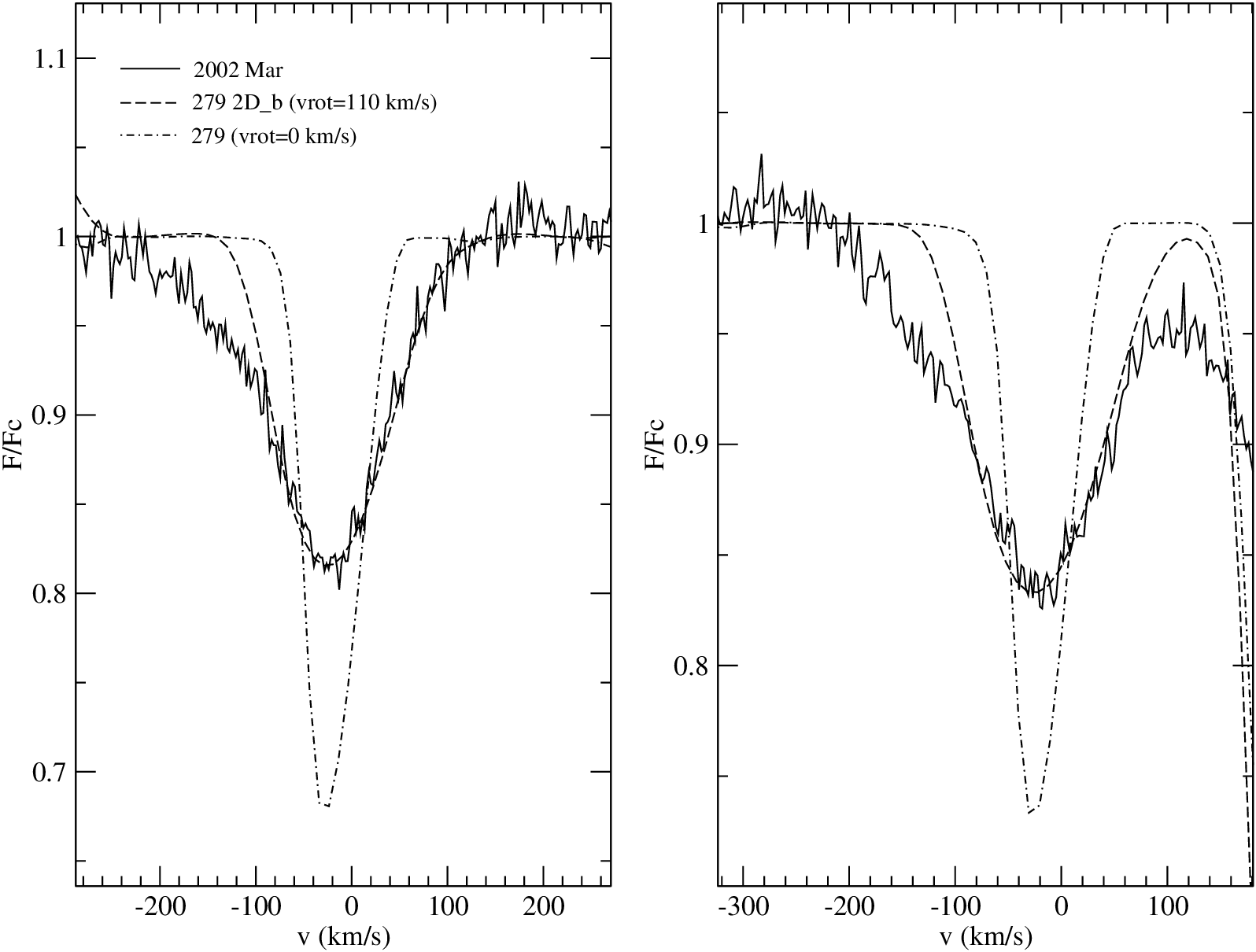}
\caption{Same as Fig. \ref{agc01}, but for the data obtained on 2002 March (full
line).
The displayed CMFGEN models were computed with (dashed) and without
(dot-dashed) a projected rotational velocity 
of 110\,km\,s$^{-1}$ .\label{agc02}}
\end{figure}

The spectra obtained during 2002 March have a S/N higher than the 2001 April dataset. In
Figure \ref{agc02} we present the best model for this epoch, compared with the
observational data. The physical parameters of the star derived from the best fits of
CMFGEN were L=1.0$\times10^{6}$\,L$_\odot$,  R$_\star$=87\,R$_{\odot}$, T${_\star}$=19,650\,K,
v$_{\infty}$=195\,km\,s$^{-1}$, $\dot{M}$=4.7$\times10^{-5}$\,M$_\odot$\,yr$^{-1}$, and f=0.25. 
The non-rotating model is inconsistent with the observed lines' width and strength.
To match the width and the shape of the lines we had to use a 2D
model with $v_{rot}$\,sin\,i=110\,km\,s$^{-1}$. The high S/N allowed us to constrain the
uncertainties to $\pm$10\,km\,s$^{-1}$. In spite of the superb fits to the line cores and red
wings, we can also see in Figure \ref{agc02} a high-velocity absorption in the blue wings,
which could not be reproduced by our models. Models with higher $\dot{M}$ could not reproduce this
feature either.

\begin{figure}
\plotone{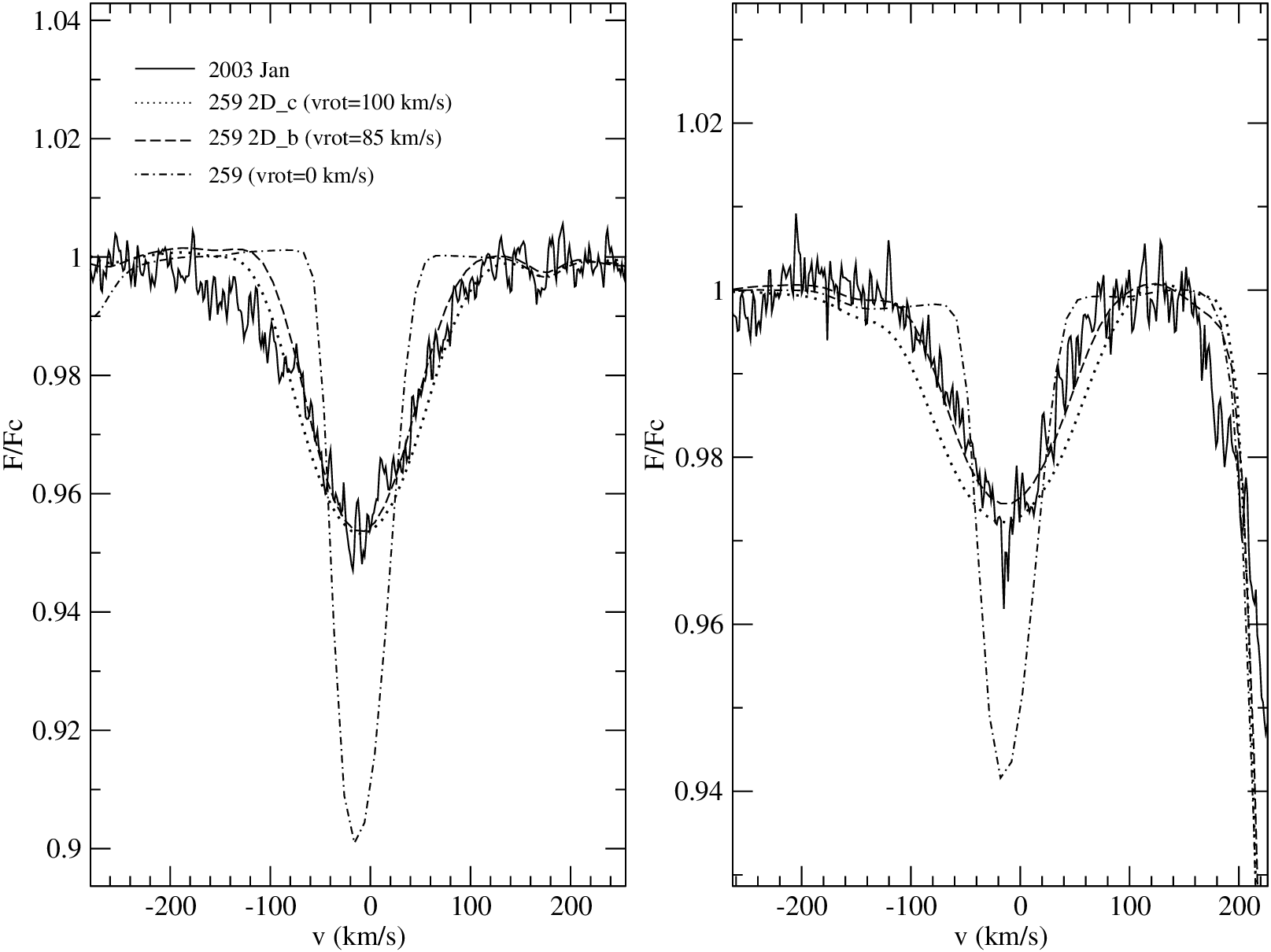}
\caption{Same as Fig. \ref{agc01}, but for the spectra gathered on 2003 January (full
line).
The CMFGEN models were computed with (dashed) and without
(dot-dashed) a projected rotational velocity
of 85\,km\,s$^{-1}$. To estimate the uncertainties, the dotted line shows the CMFGEN model with a
projected rotational velocity of
100\,km\,s$^{-1}$. \label{agc03}}
\end{figure}

We present in Figure \ref{agc03} the 2003 January observations around the \ion{Si}{4} lines compared
with our best model. At this epoch AG Car was getting colder, as confirmed by our derived stellar
temperature of 17,940 K. Thanks to the high-resolution and high S/N of the VLT data, we could still
clearly observe broader \ion{Si}{4} absorptions compared with the predictions of the non-rotating model, 
with parameters L=1.0$\times10^{6}$\,L$_\odot$, R$_\star$=103\,R$_{\odot}$,
T${_\star}$=17,940\,K, $v_{\infty}$=150\,km\,s$^{-1}$,
$\dot{M}$=6.0$\times10^{-5}$\,M$_\odot$\,yr$^{-1}$, and f=0.25. Running a 2D model with
$v_{rot}$\,sin\,i=85\,km\,s$^{-1}$, we could achieve a superb fit to the absorptions' width and
strength. A model with $v_{rot}$\,sin\,i=100\,km\,s$^{-1}$ has broader profiles than
the observed ones, especially for \ion{Si}{4} $\lambda$4116. Hence, we could constrain 
$v_{rot}$\,sin\,i=85\,$\pm$10\,km\,s$^{-1}$ in this epoch. 

In Table \ref{results} we summarize the results obtained for the 3 epochs analyzed in this Letter. The
derived luminosity agrees within 20\% with previous works \citep{sc94}, while the mass-loss rate is clearly variable.
A full discussion will be presented in a forthcoming paper. A key result is presented in the last column,
where we show the derived value of $v_{rot}\,R_\star$ . This physical parameter should be constant over
time due to angular momentum conservation, as the stellar mass is practically invariant on this timescale.
Analyzing the values derived in this work, we could verify that the rotational velocities obtained are
consistent, taking the errors into account. This reinforces our finding  that the broadening detected in
the \ion{Si}{4} photospheric lines is a consequence of rotation.

\section{\label{disc}Discussion}

Rotation is believed to play a significant role during the evolution of massive stars, especially during the
short-lived, transitional stages like the LBV phase, where the stellar and wind physical parameters change
dramatically on short timescales. Rotation  has also been invoked to explain the LBV phenomenon and the origin
of asymmetric winds found in some LBVs, like the supermassive star $\eta$ Car \citep{smith03,vb03}, and related
objects like B[e] stars \citep{zickgraf85}. For AG Car, observational evidence for wind asymmetry is
provided by polarization data \citep{leitherer94,sl94,davies05} and may have a significant impact on the
spectroscopic analysis. Such effects are currently being explored by us.

Rotation, and its effects in the wind, have to be taken into account in any theory used to explain how
the physical properties change during the S Doradus cycle for AG Car and other LBVs. One of the most
noticeable effects of a high rotation, for example, would be a significant decrease in the wind
terminal velocity \citep{fo86}. Indeed, based on our derived physical parameters, we can calculate that
during the hottest epoch (April 2001) $v_{esc}$=290\,km\,s$^{-1}$, which includes the
reduction due to the radiation pressure. The derived wind terminal velocity of 105\,km\,s$^{-1}$ is
incompatible with that surface escape velocity and the predicted value of $v_{\infty}/v_{esc}\sim1.3-2.6$ for
hot supergiants \citep{lamers95}. However, if we account for the reduction effect that a high
rotational velocity has on the effective gravity $g_{eff}$ \citep{mm_omega00}, and hence on the
surface escape velocity (defined as $v_{esc}=(2g_{eff}\,R_{\star})^{0.5}$), we would achieve
$v_{esc}\sim40-70$\,km\,s$^{-1}$ and $v_{\infty}/v_{esc}\sim1.5-2.6$.

The high rotational velocity also explains why v$_{\infty}$ was lower when the star was hotter (2001
April) compared to when the star was cooling (2002 March and 2003 January). At a first glance, this
result disagrees with the predicted temperature dependence of $v_{\infty}$ for line-driven winds
\citep{vink00}. However, the effect of rotation on reducing the wind terminal velocity is higher during the
hot phase, which can be translated to higher values of $v_{rot}/v_{crit}$, as shown in Table
\ref{results}. This can be understood in terms of the $R_{\star}^{-1}$ dependence on the rotational
velocity, while the critical rotational velocity depends on $R_{\star}^{-0.5}$. 

So far, a high rotational velocity has never been observed directly (i.\,e. in the stellar spectrum) in
any active LBV, in spite of the intense theoretical studies made in recent years. Our spectroscopic analysis
shows that AG Car, a well-known prototype of the LBV class,
rotates at a significant fraction of its break-up velocity. Interestingly, the behavior of the
rotational velocity as a function of the stellar radius during 2001-2003 is consistent with a
$R_{\star}^{-1}$ dependence (see Sec. \ref{modres}), as expected from angular momentum conservation.
Our modeling also confirms that AG Car is very luminous (L=1.0$\times10^{6}$\,L$_\odot$), and therefore
it lies close to the so-called $\Gamma\Omega$ limit \citep{mm_omega00}.  Thus, it is likely that
rotation, coupled with a high luminosity, plays a key role during the S Dor cycles, the LBV giant
eruptions and the subsequent late evolutionary stages of a massive star.

\acknowledgements 
J. H. Groh and A. Damineli thank Brazilian Agencies FAPESP (grant 02/11446-5) and
CNPq (grant 200984/2004-7) for financial support. D. J. Hillier gratefully acknowledges partial support
for this work from NASA-LTSA grant NAG5-8211. J. H. Groh thanks D. J. Hillier and the University of
Pittsburgh for hospitality and support for this work. We thank an anonymous referee for useful suggestions.

\clearpage

\begin{table} 
\caption{Summary of AG Car spectroscopic observations} 
\label{obstable}
\centering
\begin{tabular}{c c c c}
\hline\hline
Epoch & Telescope & Spectral Range ({\AA}) & R \\
\hline
2001 Apr 12 & 1.52m ESO & 3600-9200 & 48,000 \\
2002 Mar 17 & 1.52m ESO & 3600-9200 & 48,000 \\
2003 Jan 11 & 8m ESO (VLT) & 3000-10400 & 80,000 \\
\hline
\end{tabular}
\end{table} 
\clearpage

\begin{table*} 
\footnotesize
\caption{Derived physical parameters of AG Car during 2001-2003, assuming $i=90^o$} 
\label{results}
\centering
\begin{tabular}{c c c c c c c c}
\hline\hline
Epoch & T$_{\star}$ & $\dot{M}$ (10$^{-5}$\,M$_\odot$\,yr$^{-1}$) & v$_{\infty}$ (km\,s$^{-1}$) & v$_{rot}$
(km\,s$^{-1}$)& v$_{rot}$/v$_{crit}$ & R$_\star$(R$_{\odot})$ &
v$_{rot}$\,R$_\star$ (10$^4$\,R$_{\odot}$\,km\,s$^{-1}$)\tablenotemark{a} \\
\hline
2001 Apr 12 & 24,340 & 2.3 & 105 & 190 $\pm$ 30 & 0.86 & 56 & 10.7 $\pm$ 2.2\\
2002 Mar 17 & 19,650 & 4.7 & 195 & 110 $\pm$ 10 & 0.63 & 87 & 9.5 $\pm$ 1.7\\
2003 Jan 11 & 17,940 & 6.0 & 150 & 85 $\pm$ 10 & 0.57 & 103 & 9.3 $\pm$ 1.8\\
\hline
\end{tabular}
\tablenotetext{a}{We assume a conservative 10\,\% error on R$_\star$, accounting for uncertainties on
the distance, reddening law and luminosity.}
\end{table*} 
\clearpage

\end{document}